\documentclass[12 pt, onecolumn]{IEEEtran}
\usepackage{cite}
\ifCLASSINFOpdf
 \usepackage[pdftex]{graphicx}
\else
\fi
\usepackage{amsmath, amssymb, amsthm}
\usepackage{array}
\usepackage{mdwmath}
\usepackage{mdwtab}
\usepackage{tikz}
\usepackage{array}
\usepackage{url}
\newtheorem{thm}{Theorem}
\newtheorem{lemma}[thm]{Lemma}
\theoremstyle{definition}
\newtheorem{defn}[thm]{Definition}
 \newcommand{\ket}[1]{|#1\rangle}
 \DeclareMathOperator{\tr}{tr}
\DeclareMathOperator{\wt}{wt}

 \newcommand{\reflemma}[1]{Lemma~\ref{#1}}
\newcommand{\F}{\mathbb F_q^*}
\newcommand{\Hom}{\operatorname{Hom}}
 \newcommand{\Spec}{\operatorname{Spec}}
\newcommand{\orb}{\operatorname{orb}}

\begin{document}
\title{Asymmetric Quantum Codes on Toric Surfaces}
\author{Johan P. Hansen}
\thanks{Johan P. Hansen was with the Department of Mathematics, Aarhus University, Denmark e-mail: matjph@math.au.dk}

{Johan P. Hansen: Asymmetric Quantum Codes on Toric Surfaces}
\maketitle

\begin{abstract}
Asymmetric quantum error-correcting codes are quantum codes defined over biased quantum channels: qubit-flip and phase-shift errors may have equal or different probabilities. The code construction is the Calderbank-Shor-Steane construction based on two linear codes. 

We present families of toric surfaces, toric codes and associated asymmetric quantum error-correcting codes.

\end{abstract}


\IEEEpeerreviewmaketitle

\section{Introduction}

\IEEEPARstart{W}{orks} of P.W. Shor \cite{PhysRevA.52.R2493} and A.M. Steane \cite{MR1421749}, \cite{MR1398854} initiated the study and construction of quantum error-correcting codes. 
A.R. Calderbank \cite{Calderbank19961098}, 
P.W. Shor \cite{MR1450603} and 
A.M. Steane \cite{Steane19992492} produced stabilizer codes (CSS) from linear codes containing their dual codes. For details see for example \cite{959288}, \cite{MR1665774} and \cite{MR1750540}.

In  \cite{MR1749454} and \cite{MR1953195}  the author developed methods to construct linear error correcting codes from toric varieties. In \cite{MR3015727} we generalized  this to construct linear codes suitable for constructing quantum codes by the Calderbank-Shor-Steane method. Our constructions extended similar results obtained by A. Ashikhmin, S. Litsyn and M.A. Tsfasman in \cite{tsfasman} from Goppa codes on algebraic curves.

Asymmetric quantum error-correcting codes are quantum codes defined over biased quantum channels: qubit-flip and phase-shift errors may have equal or different probabilities. The code construction is the CSS construction based on two linear codes. The construction appeared originally in \cite{2007arXiv0709.3875E}, \cite{PhysRevA.75.032345} and \cite{PhysRevA.77.062335}. We present new families of toric surfaces, toric codes and associated asymmetric quantum error-correcting codes.

\hfill August 7, 2017
\subsection{Notation}
\begin{itemize}
\item $\mathbb F_q$ -- the finite field with $q$ elements of characteristic $p$.
\item $\F$ -- the invertible elements in $\mathbb F_q$.
\item $k=\overline{\mathbb F_q}$ -- an algebraic closure of $\mathbb F_q$.
\item $M \simeq \mathbb Z^2$ a free $\mathbb Z$-module of rank 2.
\item $\square \subseteq M_{\mathbb R}= M \otimes_{\mathbb Z}{\mathbb R}$ -- an integral convex polytope.
\item $X=X_{\square}$ -- the toric surface associated to the polytope $\square$.
\item $T=T_N =U_0 \subseteq X$ -- the torus.
\end{itemize}

\section{The method of toric varieties}\label{toric}
For the general theory of toric varieties we refer to \cite{MR2810322}, \cite{MR1234037} and \cite{MR922894}.

\subsection{The toric surfaces $X_b$ and their intersection theory}

Let $\mathbb F_q$ be the field with $q$ elements and let $r$ be an integer dividing $q$.
Let $b \in  \mathbb Z$ such that $0 \leq b \leq q-2$ with $a:=b+ \frac{q-2}{r} \leq q-2$.

Let $M$ be an integer lattice $M \simeq \mathbb Z^2$. Let
$N=\Hom_\mathbb Z(M,\mathbb Z)$ be the dual lattice with canonical $\mathbb Z$ -
bilinear pairing $ <\quad,\quad>: M \times N \rightarrow \mathbb Z.$ Let
$M_{\mathbb R}= M \otimes_{\mathbb Z}{\mathbb R}$ and $N_{\mathbb
R}= N\otimes_{\mathbb Z}{\mathbb R}$ with canonical  $\mathbb R$ -
bilinear pairing $ <\quad,\quad>: M_{\mathbb R} \times N_{\mathbb
R} \rightarrow  \mathbb R. $

Let  $\square_b$ in  $M_{\mathbb R}$ be the 2-dimensional integral convex polytope in $M_{\mathbb R}$ with vertices $(0,0), (a,0), (b,q-2)  \mathrm{\ and\ }  (0,q-2)$ properly contained in the square $\left[0, q- 2 \right] \times \left[0, q-2\right]$, see
Figure \ref{bpolytop}. It is the Minkowski sum of the line segment from $(0,0)$ to $(b,0)$ and the polytope $\square_0$, see Figure \ref{0polytop}.
\begin{figure}
\begin{center}
\begin{tikzpicture}[scale=0.3]
\draw[step=1cm,gray,very thin] (-1,-1) grid (16,16);
\draw[thick,dotted](0,15)--(15,15);
\draw[thick,dotted](15,0)--(15,15);
\draw[thick,dotted](0,0)--(15,0);
\draw[thick] (0,0)--(9,0)--(4,15)--(0,15)--(0,0);
\draw (-1,15) node[anchor=east] {$q-2$};
\draw[thick,->] (4,16)--(4,15.5);
\draw (4,16) node[anchor=south] {$b$};
\draw (9,-1) node[anchor=north] {$a=b+\frac{q-2}{r}$};
\draw (15,16) node[anchor=south] {$q-2$};
\draw[thick,<-] (9,-0.5)--(9,-1);
\foreach \x in {0,1,2,3,4,5,6,7,8,9,10,11,12,13,14,15} \draw (0,\x) circle (0.2cm);
\foreach \x in {0,1,2,3,4,5,6,7,8,9,10,11,12,13,14,15} \draw (1,\x) circle (0.2cm);
\foreach \x in {0,1,2,3,4,5,6,7,8,9,10,11,12,13,14,15} \draw (2,\x) circle (0.2cm);
\foreach \x in {0,1,2,3,4,5,6,7,8,9,10,11,12,13,14,15} \draw (3,\x) circle (0.2cm);
\foreach \x in {0,1,2,3,4,5,6,7,8,9,10,11,12,13,14,15} \draw (4,\x) circle (0.2cm);
\foreach \x in {0,1,2,3,4,5,6,7,8,9,10,11,12} \draw (5,\x) circle (0.2cm);
\foreach \x in {0,1,2,3,4,5,6,7,8,9} \draw (6,\x) circle (0.2cm);
\foreach \x in {0,1,2,3,4,5,6} \draw (7,\x) circle (0.2cm);
\foreach \x in {0,1,2,3} \draw (8,\x) circle (0.2cm);
\foreach \x in {0} \draw (9,\x) circle (0.2cm);
\foreach \x in {0,1,2,3,4,5,6} \draw (1,\x) circle (0.2cm);
\foreach \x in {0,1,2,3,4,5,6,7,8,9} \draw (2,\x) circle (0.2cm);
\foreach \x in {0,1,2,3,4,5,6,7,8,9,10,11,12} \draw (3,\x) circle (0.2cm);
\foreach \x in {0,1,2,3,4,5,6,7,8,9,10,11,12,13,14,15} \draw (4,\x) circle (0.2cm);
\end{tikzpicture}
\end{center}
\caption{The polytope  $\square_b$ is the polytope with
vertices $(0,0),(a=b+\frac{q-2}{r},0),(b,q-2), (0,q-2)$.} \label{bpolytop}
\end{figure}
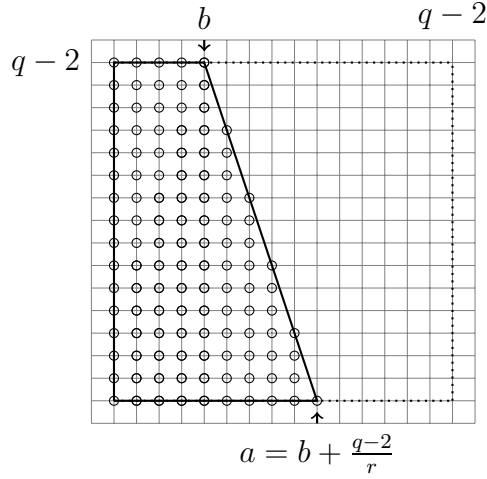
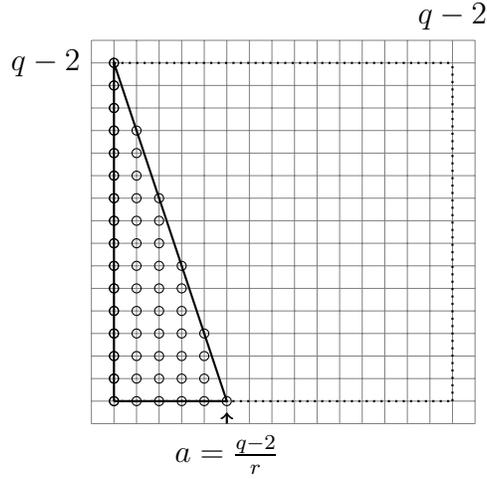
\begin{figure}
\begin{center}
\begin{tikzpicture}[scale=0.3]
\draw[step=1cm,gray,very thin] (-1,-1) grid (16,16);
\draw[thick,dotted](0,15)--(15,15);
\draw[thick,dotted](15,0)--(15,15);
\draw[thick,dotted](0,0)--(15,0);
\draw[thick] (0,0)--(5,0)--(0,15)--(0,15)--(0,0);
\draw (-1,15) node[anchor=east] {$q-2$};
\draw[thick,<-] (5,-0.5)--(5,-1);
\draw (5,-1) node[anchor=north] {$a=\frac{q-2}{r}$};
\draw (15,16) node[anchor=south] {$q-2$};
\foreach \x in {0,1,2,3,4,5,6,7,8,9,10,11,12,13,14,15} \draw (0,\x) circle (0.2cm);
\foreach \x in {0,1,2,3} \draw (4,\x) circle (0.2cm);
\foreach \x in {0} \draw (5,\x) circle (0.2cm);
\foreach \x in {0,1,2,3,4,5,6} \draw (3,\x) circle (0.2cm);
\foreach \x in {0,1,2,3,4,5,6,7,8,9} \draw (2,\x) circle (0.2cm);
\foreach \x in {0,1,2,3,4,5,6,7,8,9,10,11,12} \draw (1,\x) circle (0.2cm);
\foreach \x in {0,1,2,3,4,5,6,7,8,9,10,11,12,13,14,15} \draw (0,\x) circle (0.2cm);
\end{tikzpicture}
\end{center}
\caption{The polytope  $\square_0$ is the polytope with
vertices $(0,0),(a=\frac{q-2}{r},0),(0,q-2).$} \label{0polytop}
\end{figure}

\begin{figure}
\begin{center}
\begin{tikzpicture}[scale=0.6]
\draw[step=1cm,gray,very thin] (-5,-5) grid (5,5);
\draw (-5,0)--(5,0);
\draw (0,0)--(0,5);
\draw (0,0)--(-5,-5/3);
\draw[->, very thick] (0,0)--(0,1);
\draw[->, very thick] (0,0)--(-1,0);
\draw[->, very thick] (0,0)--(1,0);
\draw[->, very thick] (0,0)--(-3,-1);
\draw (1.8,2) node[anchor=east] {$V(\rho_2)$};
\draw (-2,1) node[anchor=north] {$V(\rho_3)$};
\draw (-4,0) node[anchor=north] {$V(\rho_4)$};
\draw (2,1) node[anchor=north] {$V(\rho_1)$};
\end{tikzpicture}
\end{center}
\caption{The refined normal fan and the 1-dimensional cones of the polytope $\square_0$ in Figure
\ref{0polytop}}\label{fan}
\end{figure}
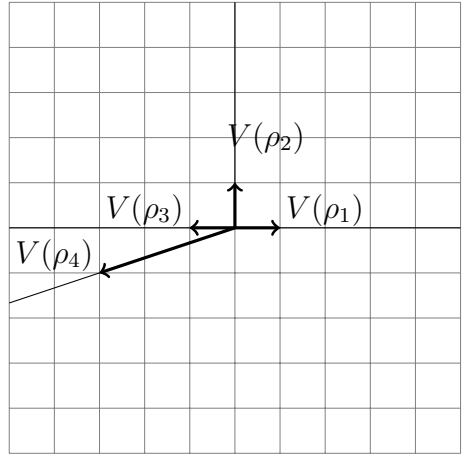

The support function 
$
h_b: N_{\mathbb R} \rightarrow \mathbb R
$
for $\square_b$ is defined as
$
h_b(n):= {\rm inf}\{<m,n> | \, m \in \square_0\}
$
and the polytope $\square_b$ can be reconstructed from the support function
\begin{equation}\label{support}
\square_b = \{m \in M |\, <m,n> \,\geq \,h(n) \quad \forall n \in N \}.
\end{equation}

The support function $h_b$ is piecewise linear in the sense
that  $N_{\mathbb R}$ is the union of
a non-empty finite collection of strongly convex polyhedral cones in $N_{\mathbb R}$ such that
$h_b$ is linear on each cone. A fan is a collection $\Delta$ of
strongly convex polyhedral cones in $N_{\mathbb R}$
such that every face of $\sigma \in \Delta $ is contained in $\Delta $ and $\sigma \cap \sigma'  \in \Delta$ for all
$\sigma , \sigma' \in \Delta $.

The {\it normal fan} $\Delta_b$ is the coarsest  fan such that $h_b$ is linear
on each $\sigma \in \Delta_b$, i.e. for all $\sigma \in \Delta_b$ there exists  $l_{\sigma} \in M$ such that
\begin{equation}
h_b(n) = <l_{\sigma},n> \quad \forall n \in \sigma.
\label{linear}
\end{equation}

Upon refinement of the normal fan, we can assume that two
successive pairs of $n(\rho)$'s generate the lattice and we obtain
{\it the refined normal fan}.

The 1-dimensional cones in the refined normal fan $\Delta_0$ of the polytope $\square_0$ are generated by unique primitive elements
$n(\rho) \in N \mathbb Cap \rho$ such that $\rho =\mathbb R_{\geq 0} n(\rho)$, specifically
are \begin{equation}
n_{\rho_1}=\begin{pmatrix}1\\0\end{pmatrix},
n_{\rho_2}=\begin{pmatrix}0\\1\end{pmatrix},
n_{\rho_3}=\begin{pmatrix}-1\\0\end{pmatrix},
n_{\rho_4}=\begin{pmatrix}-r\\-1\end{pmatrix}\ ,
\end{equation}
see Figure \ref{fan}.

There are four 2-dimensional cones in the refined normal fan $\Delta_0$:
\begin{enumerate}
\item $\sigma_1$ with faces  $\rho_1$, $\rho_2$ and $l_{\sigma_1}=\begin{pmatrix}0\\0\end{pmatrix}$
\item $\sigma_2$ with faces $\rho_2$, $\rho_3$ and $l_{\sigma_1}=\begin{pmatrix}\frac{q-2}{r}\\0\end{pmatrix}$
\item $\sigma_3$ with faces $\rho_3$, $\rho_4$ and $l_{\sigma_1}=\begin{pmatrix}\frac{q-2}{r}\\0\end{pmatrix}$
\item $\sigma_4$ with faces $\rho_4$, $\rho_1$ and $l_{\sigma_1}=\begin{pmatrix}0\\q-2\end{pmatrix}$\ .
\end{enumerate}

The 2-dimensional {\it algebraic torus} $T_N \simeq 
k^* \times k^* $ is defined by $T_N:= \Hom_\mathbb Z(M,k^*)$. The multiplicative
character $\mathbf e(m),\, m \in M$ is the homomorphism $\mathbf e(m): T \rightarrow k^*$ defined
by $\mathbf e(m)(t) = t(m)$ for $t \in T_N$. Specifically, if $\{n_1,n_2\}$ and $\{m_1,m_2\}$ are dual $\mathbb Z$-bases of $N$ and $M$ and we denote
$u_j:= \mathbf e(m_j),\,j=1,2$, then we have an isomorphism $T_N \simeq k^* \times k^* $ sending $t$ to $(u_1(t),u_2(t))$.
For $m=\lambda_1 m_1 +\lambda_2 m_2$ we have
\begin{equation}\label{e}
\mathbf e(m)(t)=u_1(t)^{\lambda_1}u_2(t)^{\lambda_2}.
\end{equation}
The {\it toric surface} $X_b$ associated to the normal fan $\Delta_b$ of $\square_b$ is
\begin{equation*}
X_b = \cup_{\sigma \in \Delta} U_{\sigma}\ ,
\end{equation*}
where $U_{\sigma}$ is the $k$-valued points of the affine scheme  $\Spec(k[\mathcal S_{\sigma}])$, i.e.,
morphisms $u : {\mathcal S}_{\sigma} \rightarrow k$ with $u(0)=1$ and $
u(m+m')= u(m)u(m')\ \forall m,m' \in \mathcal S_{\sigma}$,
where $\mathcal S_{\sigma}$ is the additive subsemigroup of $M$
\begin{equation*}
\mathcal S_{\sigma}=\{m \in M | <m,y> \geq 0 \ \forall y \in \sigma\}.
\end{equation*}

The {\it toric surface} $X_b$ is irreducible
and complete. Under the assumption that we are working with the refined normal fan,  the surface $X_b$ is non-singular.

 If $\sigma, \tau \in \Delta$ and $\tau$ is a face of
$\sigma$, then $U_{\tau}$ is an open subset of $U_{\sigma}$.
Obviously $\mathcal S_{0}=M$ and $U_{0}=T_N$ such that the algebraic
torus $T_N$ is an open subset of $X_{\square}$.

$T_N$ {\it acts algebraically} on $X_{\square}$. On $u \in U_{\sigma}$ the action of $t \in T_N$ is obtained as
\begin{equation*}
(tu)(m):=t(m)u(m) \quad  \mathrm{for\ }m \in \mathcal S_{\sigma}\ ,
\end{equation*}
such that $tu \in U_{\sigma}$ and $U_{\sigma}$ is $T_N$-stable.
The orbits of this action are in one-to-one correspondance with
$\Delta$. For each $\sigma \in \Delta$, let
$$\orb(\sigma):=\{u:M \cap \sigma \rightarrow k^* | u \text{
is a group homomorphism}\}.$$ Then $\orb(\sigma)$ is a $T_N$ orbit
in $X_{\square}$. Define $V(\sigma)$ to be the closure of
$\orb(\sigma)$ in $X_{\square}$.

A $\Delta$-linear support function $h$ gives rise to a polytope $\square$ as above and an associated Cartier
divisor 
\begin{equation*}
D_h=D_{\square}:= -\sum_{\rho \in \Delta (1)} h(n(\rho))\,V(\rho)\ ,
\end{equation*}
where $\Delta (1)$ is the 1-dimensional cones in $\Delta$.
In particular
\begin{equation*}
D_m={\rm div}(\mathbf e(-m)) \quad m \in M.
\end{equation*}
Let $h_0$ be the support function of the refined normal fan $\Delta_0$. Then 
\begin{equation}\label{Dh}
D_{h_0}= \frac{q-2}{r} V(\rho_3) + (q-2) V(\rho_4)
\end{equation}
\begin{lemma} \label{cohomology} Let  $h$ be a $\Delta$-linear support function
with associated convex polytope $\square$ and Cartier divisor $D_h=D_{\square}$. The vector space
${\rm H}^0(X,\it O_X(D_h))$ of global sections of $O_X(D_{\square})$, i.e., rational functions $f$ on
$X_{\square}$ such that
${\rm div}(f) + D_{\square} \geq 0$, has dimension $\#(M \cap \square)$ and has
$
\{\mathbf e(m) | m \in M \cap \square\}
$
as a basis.
\end{lemma}

For a $\Delta$-linear support function $h$ and a 1-dimensional
cone $\rho \in \Delta (1)$   the intersection
number $(D_h;V(\rho))$ between the Cartier divisor $D_h$ and
$V(\rho)) =\mathbb P^1$ can be calculated. The cone $\rho$ is the common face of two 2-dimensional
cones $\sigma', \sigma'' \in \Delta (2)$. Choose primitive
elements $n', n'' \in N$ such that
\begin{align*}
n'+n''& \in \mathbb R \rho\\
\sigma' + \mathbb R \rho &= \mathbb R_{\geq 0} n' + \mathbb R \rho\\
\sigma'' + \mathbb R \rho &= \mathbb R_{\geq 0} n'' + \mathbb R \rho
\end{align*}
\begin{lemma}
\label{inter}For any $l_{\rho} \in M$, such that $h$ coincides
with $l_{\rho}$ on $\rho$, let $\overline{h} = h-l_{\rho}$. Then
\begin{equation*}
(D_h;V(\rho))= -(\overline{h}(n')+\overline{h}(n'').
\end{equation*}
\end{lemma}
\begin{IEEEproof} See \cite[Lemma
2.11]{MR922894}.
\end{IEEEproof}
Let $h_0$ be the support function of the refined normal fan $\Delta_0$. Let $D_{h_0}$ be the associated Cartier divisor of (\ref{Dh}). Then $n'=n(\rho_1)$ and $n''= n(\rho_4)$ in the lemma above and $h_0=\overline{h_0}$ on $\rho_1$, therefore
\begin{equation}\label{q-2}
(D_{h_0};V(\rho_1))= -(h_0(n_{\rho_4}))=q-2\ .
\end{equation}

In the 2-dimensional non-singular case let $n(\rho)$ be a
primitive generator for the 1-dimensional cone $\rho$. There
exists an integer $c$ such that
\begin{equation*}
n'+n''+c n(\rho)=0,
\end{equation*}
$V(\rho)$ is itself a Cartier divisor and the above gives the
self-intersection number
\begin{equation*}
(V(\rho);V(\rho))=c\ .
\end{equation*}

In case of the refined normal fan $\Delta_0$  of $\square_0$, we have
\begin{equation*}
n'+n''+r n(\rho_1)=0\ ,
\end{equation*}
consequently,
\begin{equation}\label{r}
(V(\rho_1);V(\rho_1))=r.
\end{equation}

\subsection{Toric evaluation codes}\label{dimension}
\begin{defn}\label{toriccode}
For each $t \in T \simeq k^{*} \times k^{*}$, we
evaluate the rational functions in ${\rm H}^0(X_b,\it O_X(D_{\square_b}))$
\begin{eqnarray*}
    {\rm H}^0(X_b,\it O_X(D_{\square_b}))& \rightarrow & k\\
    f & \mapsto & f(t).
\end{eqnarray*}
Let ${\rm H}^0(X_b,\it O_X(D_{\square_b}))^{{\rm Frob}}$ denote the rational
functions in ${\rm H}^0(X_b,\it O_X(D_{\square_b}))$ that are invariant under
the action of the Frobenius, i.e. functions that are $\mathbb F_q$-linear
combinations of the functions $\mathbf e(m)$ in
(\ref{e}).

Evaluating in all points in $S = \F \times \F \subseteq X_b$, we obtain the code
$C_{b} \subset (\mathbb F_q)^{\# S}$ as the image
\begin{eqnarray}\label{code}
    {\rm H}^0(X_b,\it O_X(D_h))^{{\rm Frob}}& \rightarrow & C_{b} \subset (\mathbb F_q)^{\# S}
 \\
    f & \mapsto & (f(t))_{t \in T(\mathbb F_q)}
\end{eqnarray}
and the generators of the code is obtained as the image of the basis
\begin{equation*}
\mathbf e(m) \mapsto (\mathbf e(m)(t))_{t \in S}.
\end{equation*}
as in (\ref{e}).
\end{defn}

\begin{thm} \label{kode-Cb}Let $\mathbb F_q$ be the field with $q$ elements and let $r$ be an integer dividing $q$. Let $b \in  \mathbb Z$ such that $0 \leq b \leq q-2$ with $a:=b+ \frac{q-2}{r} \leq q-2$.

Let  $\square_b$ in  $M_{\mathbb R}$ be the 2-dimensional integral convex polytope in $M_{\mathbb R}$ with vertices $(0,0), (a,0), (b,q-2)  \mathrm{\ and\ }  (0,q-2)$ contained in the square $\left[0, q- 2 \right] \times \left[0, q-2\right]$, see
Figure \ref{bpolytop}. Let $C_b$ be the corresponding toric code of (\ref{code}).

Then $n:= \mathrm{length\ } C_b = (q-1)^2$ and $k=\dim C_b = \frac{1}{2} \left(\frac{q-2}{r}+1\right) q + b (q-1)$ and the minimum distance $d(C_b)=(q-1-a)(q-1)$.
\end{thm}
\begin{IEEEproof}As we evaluate in $(q-1)^2$ points on $X_b$ the length is as claimed.
The dimension $\dim C_b$ equals the number of integral points in $\Delta_b$, which is $\frac{1}{2} \left(\frac{q-2}{r}+1\right) q + b (q-1)$.

\paragraph{Minimum distance in the special case  $b=0$, see figures \ref{0polytop} and \ref{fan}}
 We bound the number of points in the support $S= \F \times \F \subseteq X_b$, where the rational functions in ${\rm H}^0(X_0,\it O_X(D_{h_0}))^{{\rm Frob}}$ evaluates to zero. 

The support $S$ is stratified by the intersections with the zeros of $\mathbf e(m_1)-\psi$, where $\psi \in \F$. A rational function $f$ can either vanish identically on a stratum or have a finite number of zeroes along the stratum.

\emph{Identically vanishing on strata:} Assume that $f$ is identically zero
along precisely $A$ of these strata. As $\mathbf
e(m_{1})-\psi$ and $\mathbf
e(m_{1})$ have the same divisors of poles, they have equivalent
divisors of zeroes, so
\begin{equation*}
(\mathbf
e(m_{1})-\psi)_{0} \sim (\mathbf
e(m_{1}))_{0}.
\end{equation*}
Therefore
\begin{equation*}
{\rm div}(f) + D_{h_0} -A (\mathbf
e(m_{1}))_{0}\geq 0
\end{equation*}
or equivalently
\begin{equation*}
f \in {\rm H}^0(X_0,\it
O_X(D_{h_0}-A(\mathbf
e(m_{1}))_{0}).
\end{equation*}
Therefore $A\leq a$ by \reflemma{cohomology}.

\emph{Vanishing in a finite number of points on a stratum:}
On any of the $q-1-A$ other strata,
the number of zeroes of $f$ is at most
the intersection number
\begin{equation}\label{intersection}
(D_{h_0}-A (\mathbf e(m_{1}))_{0};(\mathbf
e(m_{1}))_{0}) = (q-2)-A r
\end{equation}
following  (\ref{q-2}) and (\ref{r}), see  \cite{MR1866342} .

Consequently, the number of zeros is at most $A(q-1)+ (q-1-A)(q-2-Ar) \leq (q-1)^2-(q-1-a)(q-1)$ as $A \leq a$ and therefore $d(C_0) \geq (q-1-a)(q-1)$.

\paragraph{Minimum distance in the general case  $b>0$, see figure \ref{bpolytop}}
The polytope  $\square_b$  with vertices $(0,0), (a,0), (b,q-2)  \mathrm{\ and\ }  (0,q-2)$ is the Minkowski sum of the line segment from $(0,0)$ to $(b,0)$ and the polytope $\square_0$, see Figure \ref{0polytop}. Applying  \cite[Proposition 2.3]{MR2272243} and the special case $b=0$, we have the inequality $d(C_a) \geq (q-1-a)(q-1)$ also in the general case.

For pairwise different $x_1,\dots, x_a \in \F$ the function $(e(m_{1})-x_1)(e(m_{1})-x_2)\ \dots (\mathbf
e(m_{1})-x_a) \in {\rm H}^0(X_0,\it
O_X(D_{h_0}))$ vanishes in the $a(q-1)$ points $(x_i,y), i=1,\dots, a$ and $y \in \F.$
In conclusion,  we have the equality $d(C_b) = (q-1-a)(q-1)$.
\end{IEEEproof}

\subsection{Asymmetric Quantum Codes}
\subsubsection{Notation}
Let $\mathcal{H}$ be the Hilbert space
$\mathcal{H}=\mathbb C^{q^n}=\mathbb C^q \otimes \mathbb C^q \otimes ... \otimes \mathbb C^q$.
Let $\ket{x}, x \in \mathbb F_q$ be an orthonormal basis for $\mathbb C^q$.
For $a,b
\in \mathbb F_q$,  the unitary operators $X(a)$ and $Z(b)$ in $\mathbb C^q$ are
\begin{eqnarray}X(a)\ket{x}=\ket{x+a},\qquad
Z(b)\ket{x}=\omega^{\tr(bx)}\ket{x},\end{eqnarray} where
$\omega=\exp(2\pi i/p)$ is a primitive $p$th root of unity and $\tr$
is the trace operation from $\mathbb F_q$ to $\mathbb F_p$.

For $\mathbf{a}=(a_1,\dots, a_n)\in \mathbb F_q^n$ and
$\mathbf{b}=(b_1,\dots, b_n)\in \mathbb F_q^n$
\begin{eqnarray*} X(\mathbf{a}) &=& X(a_1)\otimes \dots \otimes X(a_n) \\ 
Z(\mathbf{b}) &=& Z(b_1)\otimes\ \dots \otimes
Z(b_n)\end{eqnarray*} 
are the tensor products of $n$ error operators.

With 
\begin{eqnarray*}\textbf{E}_x&=&\{X(\mathbf{a})=\bigotimes_{i=1}^n X(a_i)\mid
\mathbf{a} \in \mathbb F_q^n, a_i \in \mathbb F_q\}, \nonumber \\ \textbf{E}_z&=&\{Z(\mathbf{b})=\bigotimes_{i=1}^n Z(b_i)\mid \mathbf{b} \in
\mathbb F_q^n,b_i \in \mathbb F_q\}
\end{eqnarray*} 
the error groups $\mathbf{G}_x$ and
$\mathbf{G}_z$ are
\begin{eqnarray*} \mathbf{G}_x = \{
\omega^{c}\textbf{E}_x=\omega^{c}X(\mathbf{a})\,|\, \mathbf{a} \in
\mathbb F_q^n, c\in \mathbb F_p\},\nonumber \\
\mathbf{G}_z = \{\omega^{c}\textbf{E}_z=\omega^{c}Z(\mathbf{b})\,|\,
\mathbf{b} \in \mathbb F_{q^n} c\in \mathbb F_p\}\ .
\end{eqnarray*}

It is assumed that the groups $\mathbf{G}_x$ and  $\mathbf{G}_z$
represent the qubit-flip and  phase-shift errors.

\begin{defn}[Asymmetric quantum code]
A $q$-ary asymmetric quantum code $Q$, denoted by
$[[n,k,d_z/d_x]]_q$, is a $q^k$ dimensional subspace of the Hilbert
space $\mathbb{C}^{q^n}$ and can control all bit-flip errors up to
$\lfloor \frac{d_x-1}{2}\rfloor$ and all phase-flip errors up to
$\lfloor \frac{d_z-1}{2}\rfloor$. The code $Q$ detects $(d_x-1)$
qubit-flip errors as well as detects $(d_z-1)$ phase-shift errors.
\end{defn}

Let $C_1$ and $C_2$ be two linear error-correcting codes over the finite field $\mathbb F_q$, and let $[n,k_1,d_1]_q$ and $[n,k_2,d_2]_q$ be their parameters.
For the dual codes $C_i^{\perp}$, we have $\dim{C_i^{\perp}}=n-k_i$  and if  $C_{1}^\perp \subseteq C_{2}$ then $C_{2}^\perp \subseteq C_1$.

\begin{lemma}\label{lin/qua}
Let $C_i$ for $i=1,2$ be linear error-correcting codes with parameters $[n,k_i,d_i]_q$ such
that $C_1^\perp \subseteq C_{2}$ and $C_2^\perp \subseteq C_{1}$.
Let
$d_x= \min\big\{\wt(C_{1} \backslash C_2^\perp), \wt(C_{2}
\backslash C_{1 }^\perp)\big\}$, and $d_z= \max\big\{\wt(C_{1}
\backslash C_2^\perp), \wt(C_{2} \backslash C_{1 }^\perp)\big\}$.
Then there is an asymmetric quantum code with parameters
$[[n,k_1+k_2-n,d_z/d_x]]_q$. The quantum code is pure to its minimum
distance, meaning that if $\wt(C_1)=\wt(C_1\backslash C_2^\perp)$,
then the code is pure to $d_x$, also if $\wt(C_2)=\wt(C_2\backslash
C_1^\perp)$, then the code is pure to $d_z$.
\end{lemma}

This construction is well-known, see for
example \cite{959288}, \cite{MR1665774}, \cite{PhysRevA.52.R2493}, \cite{MR1421749}, \cite{PhysRevA.54.4741} , \cite{771249} \cite{5503199}. The error groups $\mathbf G_x$ and $\mathbf G_z$ can be mapped  to the linear codes $C_1$ and $C_2$.

\subsection{New Asymmetric Quantum Codes from Toric Codes}
Let $\mathbb F_q$ be the field with $q$ elements and let $r$ be an integer dividing $q$. 
Let $b \in  \mathbb Z$ such that $0 \leq b \leq \frac{(r-1)(q-2)}{r}$. Then the polytope $\square_b$ with vertices $(0,0), (a=b+\frac{q-2}{r}), (b,q-2), (0,q-2)$ is contained in $[0,q-2] \times [0,q-2]$. Consider the associated toric code $C_b$ of (\ref{code}). From \cite[Theorem 6]{Ruano:2009:SGT:1514450.1514783} we conclude that the dual code $C_b^{\perp}$ is the toric code associated to the polytope $\square_{b^{\perp}}$ with vertices $(0,0), (a^{\perp}=b^{\perp}+\frac{q-2}{r}), (b^{\perp},q-2), (0,q-2)$ where $b^{\perp}=\frac{(r-1)(q-2)}{r}-b$ such that $a^{\perp}=q-2-b$.

\begin{figure}
\begin{center}
\begin{tikzpicture}[scale=0.3]
\draw[step=1cm,gray,very thin] (-1,-1) grid (16,16);
\draw[thick,dotted](0,15)--(15,15);
\draw[thick,dotted](15,0)--(15,15);
\draw[thick,dotted](0,0)--(15,0);
\filldraw[thick,fill opacity=0.1 ] (0,0)--(13,0)--(8,15)--(0,15)--(0,0);
\filldraw[thick,fill opacity=0.1 ] (0,0)--(11,0)--(6,15)--(0,15)--(0,0);
\filldraw[thick,fill opacity=0.1 ] (0,0)--(9,0)--(4,15)--(0,15)--(0,0);
\filldraw[thick,fill opacity=0.1 ] (0,0)--(7,0)--(2,15)--(0,15)--(0,0);

\draw[thick,<-] (9,-0.5)--(9,-1);
\draw[thick,<-] (0,-0.5)--(0,-1);
\draw[thick,<-] (7,-0.5)--(7,-1);
\draw[thick,<-] (13,-0.5)--(13,-1);
\draw[thick,<-] (11,-0.5)--(11,-1);
\draw[thick,->] (4,16)--(4,15.5);
\draw[thick,->] (6,16)--(6,15.5);
\draw[thick,->] (8,16)--(8,15.5);
\draw[thick,->] (15,16)--(15,15.5);
\draw[thick,->] (-1,15)--(-0.5,15);
\draw[thick,->] (-1,0)--(-0.5,0);
\draw (-1,15) node[anchor=east] {$q-2$};
\draw (-1,0) node[anchor=east] {$0$};
\draw (4,16) node[anchor=south] {$b_1$};
\draw (2,16) node[anchor=south] {${b_2}^{\perp}$};
\draw (8,16) node[anchor=south] {$b_2$};
\draw (6,16) node[anchor=south] {${b_1}^{\perp}$};
\draw (0,-1) node[anchor=north] {$0$};
\draw (9,-1) node[anchor=north] {$a_1$};
\draw (13,-1) node[anchor=north] {$a_2$};
\draw (7,-1) node[anchor=north] {${a_2}^{\perp}$};
\draw (11,-1) node[anchor=north] {${a_1}^{\perp}$};
\draw (15,16) node[anchor=south] {$q-2$};
\end{tikzpicture}
\end{center}
\caption{The polytope  $\square_{b_i}$ is the polytope with
vertices $(0,0),(a_i=b_i+\frac{q-2}{r},0),(b_i,q-2), (0,q-2)$. The polytopes giving the dual toric codes have vertices $(0,0),(a_i^{\perp}=b_i^{\perp}+\frac{q-2}{r},0),(b_i^{\perp},q-2), (0,q-2)$, where $b_i^{\perp}=q-2-a_i$.}\label{inklu}
\end{figure}
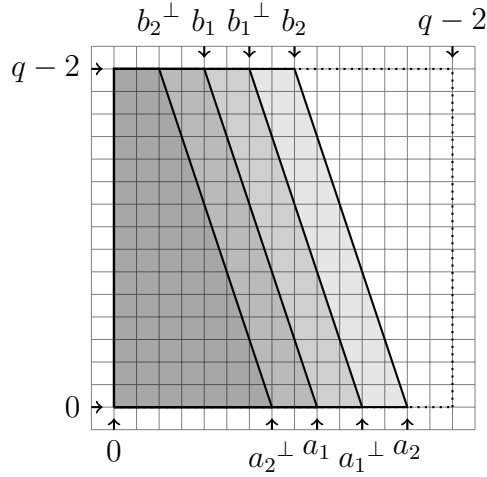

For $i=1,2$ let $b_i \in  \mathbb Z$  with $0 \leq b_i \leq \frac{(r-1)(q-2)}{r}$ and $b_1+b_2 \geq \frac{(r-1)(q-2)}{r}$. We have the inclusions of polytopes 
$\square_{b_2^{\perp}} \subseteq \square_{b_1}$ and
$\square_{b_1^{\perp}} \subseteq \square_{b_2}$, see Fig. \ref{inklu}, and corresponding inclusions of the associated toric codes of (\ref{code}):
\begin{equation*}
C_{b_2}^{\perp}=C_{b_2^{\perp}} \subseteq C_{b_1}\quad , \quad C_{b_1}^{\perp}=C_{b_1^{\perp}} \subseteq C_{b_2}\ .
\end{equation*}

The nested codes gives by the construction of Lemma \ref{lin/qua} and the discussion above rise to an asymmetric quantum code $Q_{b_1,b_2}$.

\begin{thm}[Asymmetric quatum codes $Q_{b_1,b_2}$]
Let $\mathbb F_q$ be the field with $q$ elements and let $r$ be an integer dividing $q$. For $i=1,2$ let $b_i , a_i= b_i + \frac{q-2}{r} \in  \mathbb Z$ 
$0 \leq b_i \leq \frac{(r-1)(q-2)}{r}$ and $b_1+b_2 \geq \frac{(r-1)(q-2)}{r}$. 

Then there is an asymmetric quantum code $Q_{b_1,b_2}$ with parameters
$[[(q-1)^2, \frac{1}{2} \left(\frac{q-2}{r}+1\right) q + (b_1+b_2) (q-1),d_z/d_x]]_q$, where
\begin{eqnarray*}
d_z=(q-1-\min\{b_1,b_2\})(q-1)\\
d_x=(q-1-\max\{b_1,b_2\})(q-1)
\end{eqnarray*}
 If $b_1+b_2 \neq \frac{(r-1)(q-2)}{r}$ the quantum code is pure to $d_x$ and $d_z$.
 \end{thm}
\begin{IEEEproof} The parameters and claims follow directly from Lemma \ref{lin/qua} and Theorem \ref{kode-Cb}.
\end{IEEEproof}


\begin{thebibliography}{10}

\bibitem{5503199}
S.~A. Aly and A.~Ashikhmin, \emph{Nonbinary quantum cyclic and subsystem codes
  over asymmetrically-decohered quantum channels}, 2010 IEEE Information Theory
  Workshop on Information Theory (ITW 2010, Cairo), Jan 2010, pp.~1--5.

\bibitem{959288}
A.~Ashikhmin and E.~Knill, \emph{Nonbinary quantum stabilizer codes}, IEEE
  Transactions on Information Theory \textbf{47} (2001), no.~7, 3065--3072.

\bibitem{tsfasman}
A.~Ashikhmin, S.~Litsyn, and M.A. Tsfasman, \emph{Asymptotically good quantum
  codes}, Physical Review A - Atomic, Molecular, and Optical Physics
  \textbf{63} (2001), no.~3, 1--5.

\bibitem{MR1665774}
A.~Robert Calderbank, Eric~M. Rains, P.~W. Shor, and Neil J.~A. Sloane,
  \emph{Quantum error correction via codes over {${\rm GF}(4)$}}, IEEE Trans.
  Inform. Theory \textbf{44} (1998), no.~4, 1369--1387. \MR{1665774}

\bibitem{Calderbank19961098}
A.R. Calderbank and P.W. Shor, \emph{Good quantum error-correcting codes
  exist}, Physical Review A - Atomic, Molecular, and Optical Physics
  \textbf{54} (1996), no.~2, 1098--1105.

\bibitem{MR2810322}
David~A. Cox, John~B. Little, and Henry~K. Schenck, \emph{Toric varieties},
  Graduate Studies in Mathematics, vol. 124, American Mathematical Society,
  Providence, RI, 2011. \MR{2810322}

\bibitem{2007arXiv0709.3875E}
Z.~W.~E. {Evans}, A.~M. {Stephens}, J.~H. {Cole}, and L.~C.~L. {Hollenberg},
  \emph{{Error correction optimisation in the presence of X/Z asymmetry}},
  ArXiv e-prints (2007).

\bibitem{MR1234037}
William Fulton, \emph{Introduction to toric varieties}, Annals of Mathematics
  Studies, vol. 131, Princeton University Press, Princeton, NJ, 1993, The
  William H. Roever Lectures in Geometry. \MR{1234037 (94g:14028)}

\bibitem{MR1749454}
Johan~P. Hansen, \emph{Toric surfaces and error-correcting codes}, Coding
  theory, cryptography and related areas ({G}uanajuato, 1998), Springer,
  Berlin, 2000, pp.~132--142. \MR{1749454 (2000m:94036)}

\bibitem{MR1953195}
\bysame, \emph{Toric varieties {H}irzebruch surfaces and error-correcting
  codes}, Appl. Algebra Engrg. Comm. Comput. \textbf{13} (2002), no.~4,
  289--300.

\bibitem{MR3015727}
\bysame, \emph{Quantum codes from toric surfaces}, IEEE Trans. Inform. Theory
  \textbf{59} (2013), no.~2, 1188--1192. \MR{3015727}

\bibitem{MR1866342}
S{\o}ren~Have Hansen, \emph{Error-correcting codes from higher-dimensional
  varieties}, Finite Fields Appl. \textbf{7} (2001), no.~4, 531--552.
  \MR{1866342 (2003d:94121)}

\bibitem{PhysRevA.75.032345}
Lev Ioffe and Marc M\'ezard, \emph{Asymmetric quantum error-correcting codes},
  Phys. Rev. A \textbf{75} (2007), 032345.

\bibitem{MR2272243}
John Little and Hal Schenck, \emph{Toric surface codes and {M}inkowski sums},
  SIAM J. Discrete Math. \textbf{20} (2006), no.~4, 999--1014 (electronic).
  \MR{2272243}

\bibitem{MR922894}
Tadao Oda, \emph{Convex bodies and algebraic geometry}, Ergebnisse der
  Mathematik und ihrer Grenzgebiete (3) [Results in Mathematics and Related
  Areas (3)], vol.~15, Springer-Verlag, Berlin, 1988, An introduction to the
  theory of toric varieties, Translated from the Japanese. \MR{922894
  (88m:14038)}

\bibitem{Ruano:2009:SGT:1514450.1514783}
Diego Ruano, \emph{On the structure of generalized toric codes}, J. Symb.
  Comput. \textbf{44} (2009), no.~5, 499--506.

\bibitem{PhysRevA.52.R2493}
Peter~W. Shor, \emph{Scheme for reducing decoherence in quantum computer
  memory}, Phys. Rev. A \textbf{52} (1995), R2493--R2496.

\bibitem{MR1450603}
\bysame, \emph{Fault-tolerant quantum computation}, 37th {A}nnual {S}ymposium
  on {F}oundations of {C}omputer {S}cience ({B}urlington, {VT}, 1996), IEEE
  Comput. Soc. Press, Los Alamitos, CA, 1996, pp.~56--65. \MR{1450603}

\bibitem{MR1398854}
A.~M. Steane, \emph{Error correcting codes in quantum theory}, Phys. Rev. Lett.
  \textbf{77} (1996), no.~5, 793--797. \MR{1398854}

\bibitem{PhysRevA.54.4741}
\bysame, \emph{Simple quantum error-correcting codes}, Phys. Rev. A \textbf{54}
  (1996), 4741--4751.

\bibitem{771249}
A.~M. Steane, \emph{Quantum reed-muller codes}, IEEE Transactions on
  Information Theory \textbf{45} (1999), no.~5, 1701--1703.

\bibitem{Steane19992492}
A.M. Steane, \emph{Enlargement of calderbank-shor-steane quantum codes}, IEEE
  Transactions on Information Theory \textbf{45} (1999), no.~7, 2492--2495.

\bibitem{MR1421749}
Andrew Steane, \emph{Multiple-particle interference and quantum error
  correction}, Proc. Roy. Soc. London Ser. A \textbf{452} (1996), no.~1954,
  2551--2577. \MR{1421749}

\bibitem{MR1750540}
Andrew~M. Steane, \emph{Quantum error correction}, Introduction to quantum
  computation and information, World Sci. Publ., River Edge, NJ, 1998,
  pp.~184--212. \MR{1750540}

\bibitem{PhysRevA.77.062335}
Ashley~M. Stephens, Zachary W.~E. Evans, Simon~J. Devitt, and Lloyd C.~L.
  Hollenberg, \emph{Asymmetric quantum error correction via code conversion},
  Phys. Rev. A \textbf{77} (2008), 062335.

\end{thebibliography}

\providecommand{\bysame}{\leavevmode\hbox to3em{\hrulefill}\thinspace}
\providecommand{\MR}{\relax\ifhmode\unskip\space\fi MR }
\providecommand{\MRhref}[2]{%
  \href{http://www.ams.org/mathscinet-getitem?mr=#1}{#2}
}
\providecommand{\href}[2]{#2}

\end{document}